\pdfoutput=1
\newif\ifarxiv

\arxivtrue

\documentclass[10pt,journal,compsoc]{IEEEtran}
\usepackage{microtype}

\ifCLASSOPTIONcompsoc
  \usepackage[nocompress]{cite}
\else
  \usepackage{cite}
\fi
\usepackage{tikz}
\usepackage{tikzscale}
\usetikzlibrary{arrows}
\usepackage{xcolor}
\usepackage{nicefrac}
\usepackage{bm}
\usepackage{amsfonts}
\usepackage{textcomp}
\usepackage{pgfplots}
\pgfplotsset{compat=newest}
\usepgfplotslibrary{groupplots}
\usepgfplotslibrary{dateplot}
\usetikzlibrary{pgfplots.groupplots}

\newcommand\CHANGED[1]{#1}

\ifCLASSINFOpdf
\else
\fi
\usepackage{amsmath}
\ifCLASSOPTIONcompsoc
  \usepackage[caption=false,font=footnotesize,labelfont=sf,textfont=sf]{subfig}
\else
  \usepackage[caption=false,font=footnotesize]{subfig}
\fi
\usepackage{url}

\begin{document}
\urlstyle{tt}
%
\title{A Comment on Privacy-Preserving Scalar Product Protocols as proposed in "SPOC"}
%
%
%
%

\author{Thomas Schneider and
		Amos Treiber
\IEEEcompsocitemizethanks{\IEEEcompsocthanksitem Thomas Schneider and Amos Treiber are members of the Cryptography and Privacy Engineering Group (ENCRYPTO) at TU Darmstadt, Germany. E-Mail: \{schneider, treiber\}@encrypto.cs.tu-darmstadt.de
}
\ifarxiv\thanks{\textcopyright 2019 IEEE. Personal use of this material is permitted. Permission from IEEE must be obtained for all other uses, including reprinting/republishing this material for advertising or promotional purposes, collecting new collected works for resale or redistribution to servers or lists, or reuse of any copyrighted component of this work in other works.}
\fi
}
\IEEEtitleabstractindextext{%
\begin{abstract}
Privacy-preserving scalar product (PPSP) protocols are an important building block for secure computation tasks in various applications. Lu et al. (TPDS'13) introduced a PPSP protocol that does not rely on cryptographic assumptions and that is used in a wide range of publications to date. In this comment paper, we show that Lu et al.'s protocol is insecure and should not be used. We describe specific attacks against it and, using impossibility results of Impagliazzo and Rudich (STOC'89), show that it is inherently insecure and cannot be fixed without \CHANGED{relying on at least some cryptographic assumptions}.
\end{abstract}

}

\maketitle

\IEEEdisplaynontitleabstractindextext

%
\IEEEpeerreviewmaketitle

\IEEEraisesectionheading{\section{Introduction}\label{sec:introduction}}
\IEEEPARstart{T}{he scalar} product is a fundamental operation in linear algebra that is used in a variety of fields, e.g., serving as the basis of deep neural networks, biometric characterization, or computer graphics. Suppose two parties $P_0$ and $P_1$ with respective input vectors $\vec{a}$ and $\vec{b}$ want to \CHANGED{securely} compute the scalar product $\vec{a} \cdot \vec{b} = \sum_{i=1}^{n} a_ib_i$ such that $P_0$ obtains the result $\vec{a} \cdot \vec{b}$ without revealing anything else about $\vec{b}$ to $P_0$ or anything about $\vec{a}$ to $P_1$. This \emph{secure \CHANGED{two-party} computation} of the scalar product is an important building block for preserving privacy in many applications. In 2013, Lu et al.~\cite{lu2013spoc} proposed a privacy-preserving scalar product (PPSP) protocol in their paper titled ``SPOC: A Secure and Privacy-Preserving Opportunistic Computing Framework for Mobile-Healthcare Emergency''. This protocol relies on ``multi-party random masking and polynomial aggregation techniques''~\cite{lu2014toward}, where \CHANGED{absolutely }no public-key cryptography is used. \CHANGED{In fact, their protocol does not make any cryptographic assumptions at all and the authors claim that it achieves information-theoretic security. As shown in~\cite{lu2014toward}}, the protocol is much faster than public-key based protocols using homomorphic encryption. Since then, this protocol has been and is still used in many privacy-preserving solutions\CHANGED{, e.g.,~\cite{huang2016fssr,kaur2019clampp,liu2018efficient,luan2015social,luo2015nmhp,rahulamathavan2015hide,rahulamathavan2017efficientARANK,rahulamathavan2018redesign,rahulamathavan2018privacy,wang2018efficient,wang2015pguide,wang2015lip3,yang2016sfpm,zhu2018achieve,zhu2016epcs,zhu2017efficientARANK,zhu2017efficient, zhu2014efficient},} including 
support vector machines~\cite{zhu2016epcs}, 
facial expression classification~\cite{rahulamathavan2017efficientARANK}, medical pre-diagnosis~\cite{zhu2017efficientARANK}, and speaker verification~\cite{rahulamathavan2018redesign,rahulamathavan2018privacy}.

In this comment paper, we present \CHANGED{devastating }attacks against the original~\cite{lu2013spoc} and subsequent~\cite{lu2014toward} versions of Lu et al.'s protocol. Our attacks fully break privacy and show that the protocol should not be used in applications. Before presenting our concrete attacks \CHANGED{in~\S\ref{sec:flaws}}, we first show \CHANGED{in~\S\ref{sec:seccomp} }why Lu et al.'s protocol is \emph{inherently} insecure and can only be fixed if \CHANGED{at least some }public-key cryptography \CHANGED{is} used.

\section{Lu et al.'s Protocol Cannot Be Secure}\label{sec:seccomp}
\CHANGED{A fundamental issue with privacy-preserving tasks is that the absence of attacks does not guarantee privacy. To assure the privacy of new protocols, a formal proof of security is needed. Using established simulation-based security notions, such proofs show that \emph{only} what can be computed from a priori information can be learned by executing the protocol. In this section, we will show that Lu et al.'s protocol cannot be secure under the established security notions.

\subsection{Formalizing Secure Two-Party Computation}\label{sec:seccomp:formulation}}
Formally, the secure \CHANGED{two-party} computation \CHANGED{(STPC) of a function~$f(a,b)$ on inputs~$a$ from~$P_0$ and~$b$} from~$P_1$ by a protocol~$\Pi$ is defined by \CHANGED{a \emph{simulator} $S=(S_1, S_2)$ that~\emph{simulates}} the \CHANGED{\emph{views} of the parties participating in}~$\Pi$~\cite[chapter~7]{goldreich2009foundations}:
\CHANGED{\begin{equation*}
	\begin{split}
	\{S_0(a, f(a,b))\}_{a,b} &\stackrel{c}{\approx} \{\text{view}^\Pi_0(a,b)\}_{a,b}\text{ and}\\
	\{S_1(b, f(a,b))\}_{a,b} &\stackrel{c}{\approx} \{\text{view}^\Pi_1(a,b)\}_{a,b},
	\end{split}
\end{equation*}}where $\stackrel{c}{\approx}$ denotes \emph{computational indistinguishability}. \CHANGED{$S_{i\in\{0,1\}}$ is computationally (ploynomial-time) bounded and needs to simulate $\text{view}^\Pi_i$, which contains all incoming messages received by $P_i$ during the execution of~$\Pi$. If such a simulator exists, then the protocol is considered secure because everything that can be learned from participating in the protocol ($\{\text{view}^\Pi_i(a,b)\}_{a,b}$) can also be learned by information that is known to the party anyway ($S_i$ sees either the input $a$ or $b$ and the output $f(a,b)$). Conversely, if no such simulator exists, then the distribution generated by \emph{any} $S$ can be distinguished from the distribution of the views of the protocol execution, meaning that a party observing the view reveals more information than just knowing its own input and the output. This is the established notion and the de facto standard to model secure computation tasks for privacy-preserving solutions. Thus,} in order to ensure the security of a protocol, a security proof of indistinguishability is needed~\cite{lindell2017simulate}. \CHANGED{The model we are concerned with here is in the context of \emph{semi-honest} (or \emph{passive}) security, where $P_0$ and $P_1$ honestly follow the protocol but try to learn additional information.}

\CHANGED{In the above definition, it suffices to show that one party can distinguish between different inputs of the other party based on the observed execution of a protocol to break its privacy. For instance, in our specific attacks against Lu et al.'s privacy-preserving protocol (cf.~\CHANGED{\S\ref{sec:flaws:attacks}}), we will show that~$P_0$ can distinguish between different inputs of~$P_1$ regardless of the output, thereby learning more than the minimal amount of information implied by the input and the output. Because this additional information is hard to specify and highly depends on the use case, protocols where this distinction is possible are considered insecure.

\subsection{A secure PPSP Protocol has to rely on Cryptographic Hardness Assumptions}\label{sec:seccomp:impossiblity} }

\begin{figure}[t]
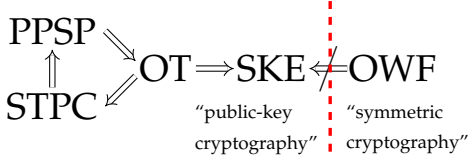

	\centering
	\ifarxiv
	\includegraphics[width=0.35\textwidth]{crypto.tikz}
	\else
	\includegraphics[width=0.35\textwidth]{gfx/crypto.tikz}
	\fi
	\caption{\CHANGED{Relations of privacy-preserving scalar product (PPSP), oblivious transfer (OT), secure two-party computation (STPC), symmetric key exchange (SKE), and one-way functions (OWF). The black box separation between ``public-key'' and ``symmetric cryptography'' shows that SKE cannot be based on OWF~\cite{impagliazzo1989limits}. Therefore, a PPSP protocol has to be based on public-key cryptographic assumptions.}}
	\label{fig_crypto}
\end{figure}

\CHANGED{In the following, we will put Lu et al.'s PPSP protocol in relation to well-established cryptographic primitives, showing that PPSP has to rely on public-key cryptography. A summary of these relations can be found in Figure~\ref{fig_crypto}.

More precisely, PPSP is closely related to a primitive called oblivious transfer (OT)}. In OT, a party $P_0$ inputs a choice bit $b$ and $P_1$ inputs two bits $(x_0,x_1)$. $P_0$ receives $x_b$ as output without learning any information about $x_{1-b}$ and without revealing any information about $b$ to $P_1$. \CHANGED{OT is a strong primitive that implies many more fundamental cryptographic building blocks such as STPC~\cite{kilian1988founding}.

Of course, STPC can be used to realize PPSP and known and \emph{secure} PPSP protocols usually rely on STPC based on homomorphic encryption or OT~\cite{demmler2015}. Since OT implies STPC, it follows that OT implies PPSP. Conversely, the existence of a PPSP protocol would imply OT, as OT is just a special case of PPSP where $\vec{a} = (\overline{b},b)$ and $\vec{b}=(x_0,x_1)$. Therefore, PPSP is equivalent to OT and requires the same assumptions required for OT like, e.g., public-key cryptography, noisy channels, or hardware tokens. 

OT can also be used to implement symmetric key exchange~(SKE)~\cite{BlumKA, RabinKA}. }Impagliazzo and Rudich~\cite{impagliazzo1989limits} proved that a black-box reduction of \CHANGED{SKE} to one-way functions \CHANGED{(the central building block of symmetric cryptography)} would imply $P\neq NP$. This means that \CHANGED{SKE and thus OT \emph{very likely} require at least some complexity-theoretic assumptions of public-key cryptography, as otherwise a proof of $P\neq NP$ would be found. As such, all PPSP protocols that rely solely on symmetric cryptography or make no cryptographic hardness assumptions at all (like Lu et al.'s protocol) must be flawed.}

\section{Lu et al.'s Protocol Is Insecure}\label{sec:flaws}

\begin{figure}[t]
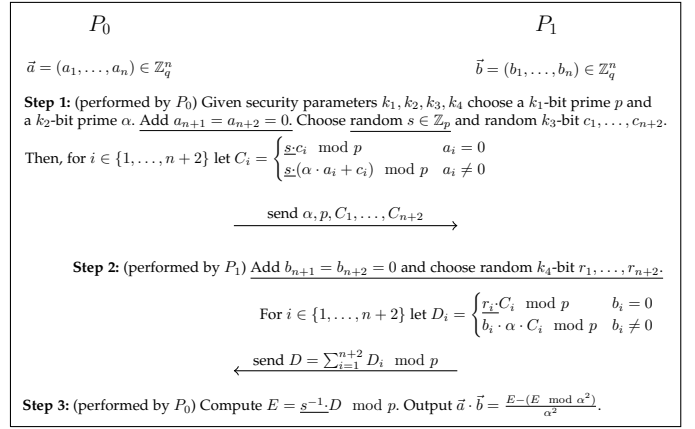

	\centering
	\ifarxiv
		\includegraphics[width=0.49\textwidth]{protocol.tikz}
	\else
		\includegraphics[width=0.49\textwidth]{gfx/protocol.tikz}
	\fi
	\caption{Lu et al.'s PPSP protocol~\cite{lu2013spoc} with the extensions of~\cite{lu2014toward} (underlined).}
	\label{fig_protocol}
\end{figure}

Lu et al.'s PPSP protocol first appeared in~\cite{lu2013spoc} as a sub-protocol in a privacy-preserving healthcare framework and was later extended in~\cite{lu2014toward} by introducing fixes to preserve privacy. The protocol is shown in Figure~\ref{fig_protocol}, with the extensions of~\cite{lu2014toward} underlined. Before presenting our specific attacks, we briefly outline how the protocol works. 

\CHANGED{
\subsection{How the Protocol is supposed to work}\label{sec:flaws:correctness}}

Correctness stems from \CHANGED{the observation that} $E = \sum_{a_i\neq 0, b_i\neq 0}a_ib_i \alpha^2 + \sum_{a_i= 0, b_i\neq 0} b_ic_i\alpha + \sum_{a_i \neq 0, b_i= 0} r_i(a_i\alpha + c_i) + \sum_{a_i= 0, b_i= 0}r_ic_i$ and therefore~${E \mod \alpha^2}$ contains all addends that are not multiples of $\alpha^2$, i.e., all addends except $\sum_{a_i\neq 0, b_i\neq 0}a_ib_i \alpha^2$. Thus, $\vec{a} \cdot \vec{b} = \frac{E - (E \mod \alpha^2)}{\alpha^2} $ under the constraint that $\sum_{a_i\neq 0, b_i\neq 0}a_ib_i \alpha^2 + \sum_{a_i= 0, b_i\neq 0} b_ic_i\alpha + \sum_{a_i \neq 0, b_i= 0} r_i(a_i\alpha + c_i) + \sum_{a_i= 0, b_i= 0} r_ic_i< p$ and $\sum_{a_i= 0, b_i\neq 0} b_ic_i\alpha + \sum_{a_i \neq 0, b_i= 0} r_i(a_i\alpha + c_i) + \sum_{a_i= 0, b_i= 0} r_ic_i< \alpha^2$. 
\CHANGED{To make the analysis of our attacks easier, we translate the latter inequality onto the corresponding bit-length parameters, resulting in the following conditions \emph{necessary} for correctness:
\begin{subequations}\label{eq:constraints}
		\begin{align}
	\log_2 n + \log_2 q + k_3 &< k_2 \label{eq:constraints4},\\
	\log_2 n + \log_2 q + k_4 & < k_2 \label{eq:constraints5},\\
	\log_2n + k_3 + k_4 &< 2k_2 \label{eq:constraints6}.
	\end{align}
\end{subequations}

A violation of any of the above inequalities would result in the protocol being incorrect for some or even all inputs. The parameters used for randomly masking the inputs,~$k_3$ and $k_4$, are both set to $128$ to allow for a randomness source of 128-bit. As a result of the above constraints when assuming an input space of~${n=q=2^{32}}$, the parameters are set to~$k_1 = 512$ and~$k_2=200$ in~\cite{lu2014toward} to ensure correctness.} Similar assumptions can be found in the original protocol~\cite{lu2013spoc}.

\CHANGED{The protocol's security is entirely based on masking values with random addends or factors. In the first step,~$P_0$ masks all values $C_i$ by multiplying with $s$. For $a_i=0$, just a random $c_i$ is masked, while a random $c_i$ added to~$\alpha \cdot a_i$ is masked otherwise. The intention behind $s$ and all $c_i$ is to hide any information about $a_i$ and, indeed, it is impossible to distinguish between different $a_i$ based on the uniform distribution from which all $c_i$ are drawn. $\alpha$ and $p$ serve no security purpose but ensure correctness. In step 2, $P_1$ either randomizes $C_i$ by multiplying with the random $r_i$ or it just multiplies $b_i\cdot \alpha$ to $C_i$. The supposed idea here is that, because~$b_{n+1}=b_{n+2}=0$, $\sum_{i} C_i$ is randomized by the addends~$r_{n+1}\cdot C_{n+1}$ and~$r_{n+2}\cdot C_{n+2}$. Thus it \emph{seems} that different values of $D$ from different $\vec{b}$ should not be distinguishable. There exist some proof sketches of the protocols in~\cite{lu2013spoc,lu2014toward} and some of the works building on them. The security analyses do not rely on the established indistinguishability-based security notions presented in~\S\ref{sec:seccomp:formulation}, but instead make use of ad-hoc security notions that are based around the principle that the input cannot be reconstructed. Below, we will present specific distinguishing attacks that even allow $P_0$ to \emph{check} whether $P_1$'s input is a candidate $\vec{b}$. This obviously violates privacy and shows that contrary to the established primitives, the ad-hoc security definitions used for the proofs do not capture any useful sense of privacy.

\subsection{Our Specific Attacks}\label{sec:flaws:attacks}}

One can immediately see why the \CHANGED{original} protocol of~\cite{lu2013spoc} is broken: $D = \sum_{b_i=0}C_i + \sum_{b_i\neq 0} b_i\cdot \alpha \cdot C_i$. Since $D$ is completely deterministic and depends only on $\alpha, C_i,$ and $\vec{b}$, party $P_0$ can easily distinguish different values of $\vec{b}$ because it knows $\alpha$ and all $C_i$. For instance, for $\vec{b}=\vec{0}$, $P_1$ will return $\sum_{i=1}^{n} C_i$ whereas for $\vec{b'} = (1, 0, \ldots, 0)$, $P_1$ will return $\alpha\cdot C_1 + \sum_{i=2}^{n} C_i$. This attack works for any value of $\vec{a}$.

\CHANGED{\subsubsection{Attack on the fixed Protocol for $\vec{a}=\vec{0}$}\label{sec:flaws:attacks:subs1}

The above }vulnerability was fixed in~\cite{lu2014toward} by introducing random addends to $D$ via $b_{n+1}=b_{n+2}=0$. Operations based on public-key cryptography still do not appear in the protocol. Thus, the security of this version is implausible as well (cf.~\CHANGED{\S\ref{sec:seccomp:impossiblity}}). Indeed, we found another attack that can distinguish different~$\vec{b}$. \CHANGED{At first, } we consider this attack for the case of $\vec{a} = \vec{0}$, because then the output of the ideal functionality is equal to $0$ and yields no knowledge about $\vec{b}$. \CHANGED{In that case, the ability to distinguish any distinct $\vec{b}$ clearly demonstrates that information about the inputs is leaked.} 
Using the following strategy, $P_0$ can distinguish between $\vec{b}=\vec{0}$ and $\vec{b'}=(1, 0, \ldots, 0)$ after computing $E$ in step 3:
\CHANGED{\begin{equation*}
	\text{If } \nicefrac{E}{\alpha} \approx c_1 \text{then output guess }\vec{b'}\text{, else output guess }\vec{b}.
\end{equation*}

Our attack relies on the different sizes of the parameters and works because they create a conflict between security and correctness: to prevent our attack, the parameters have to be changed in such a way that they violate the correctness constraints. More specifically, for~$\vec{b'}$,~$P_0$ receives $E = \alpha \cdot c_1 + \sum_{i=2}^{n+2} r_ic_i$. In our attack, $P_0$ will compute $\nicefrac{E}{\alpha} = c_1 + \frac{\sum_i r_ic_i}{\alpha}$. Since $|c_1| = k_3$ and $|\frac{\sum_i r_ic_i}{\alpha}| = \log_2 n + k_3 + k_4 - k_2$,  $\nicefrac{E}{\alpha} \approx c_1$ except for the $\log_2 n + k_3 + k_4 - k_2 $ least significant bits. Conversely, for~$\vec{b}$,~$P_0$ receives $E = \sum_{i=1}^{n+2} r_ic_i$ and thus will only obtain some $|\nicefrac{E}{\alpha}| = \log_2 n + k_3 + k_4 - k_2$ bit integer. Hence, to make our distinction impossible, the parameters need to satisfy \emph{at least} $k_3 \leq \log_2 n + k_3 + k_4 - k_2 \Leftrightarrow k_2 - \log_2 n \leq k_4$, which violates Equation~\ref{eq:constraints5} necessary for correctness.

The attack can also be extended to distinguish between \emph{any} $\vec{b}$ by checking whether $\nicefrac{E}{\alpha} \approx \sum_{i}b_i c_i$. Similar to the reasoning above, the attack can only be prevented if at least $k_2 \leq k_4$, which also violates Equation~\ref{eq:constraints5}. Not only does this break privacy because it allows for distinguishing any $\vec{b}$, this also enables an adversary to check whether a suspected input $\vec{b}$ is the real one.

\subsubsection{Attack on the fixed Protocol for any $\vec{a}$}\label{sec:flaws:attacks:subs2}

Even though the previous attack is enough to violate privacy, we will further show how to adapt it when using any~$\vec{a}$ as input. Knowing its own input~$\vec{a}$ and the suspected input~$\vec{b}$,~$P_0$ just checks whether $\nicefrac{E}{\alpha} \approx \sum_{a_i\neq 0} a_ib_i\alpha + \sum_{a_i=0}b_i c_i$. Analogously to the analysis in \S\ref{sec:flaws:attacks:subs1}, this distinction could only fail if $|\sum_{a_i\neq 0} a_ib_i\alpha + \sum_{a_i = 0}b_ic_i| \leq |\sum_{a_i \neq 0, b_i= 0} r_ic_i + \sum_{a_i= 0, b_i= 0} \frac{r_ic_i}{\alpha}|$ which, taking into account Equations~\ref{eq:constraints4} and~\ref{eq:constraints5}, requires that $2(k_2 + \log_2 q) - k_3 \leq k_4$. This would contradict Equation~\ref{eq:constraints6} and therefore violate correctness.}

\CHANGED{\subsubsection{Evaluation}\label{sec:flaws:attacks:eval}}

To show the feasibility of our attacks, we implemented them alongside the protocol. \CHANGED{Our implementation shows that for the parameters used in~\cite{lu2014toward}, any user input $\vec{b}$ can easily be distinguished and even detected }by $P_0$. The implementation is freely available as open source and can be found online at~\url{https://encrypto.de/code/SPOCattack}.

\begin{figure}
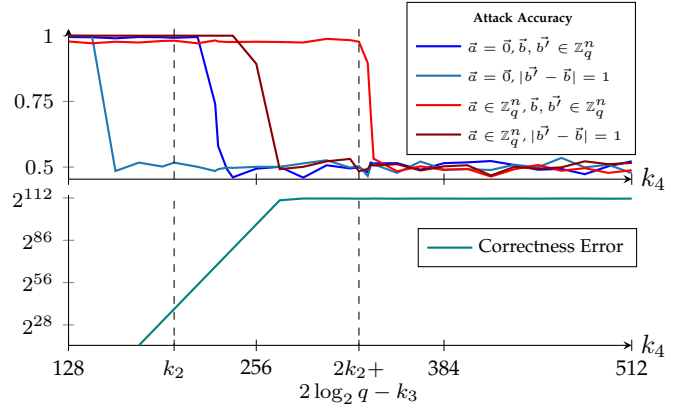

	\centering
	\ifarxiv
		\includegraphics[width=0.49\textwidth]{protocol_eval_plot.tikz}
	\else
		\includegraphics[width=0.49\textwidth]{gfx/protocol_eval_plot.tikz}
	\fi
	\caption{\CHANGED{Correctness of Lu et al.'s protocol~\cite{lu2014toward} (absolute error) for $\vec{b}=\vec{0}$ and accuracy of our attacks in distinguishing two distinct $\vec{b},\vec{b'}$, given for $n=256$, $q=2^{32}$, $k_1=512, k_2=200, k_3=128$, and varying $k_4$. Vectors are created uniformly at random, unless indicated otherwise. As predicted, the accuracies of our attacks against random $\vec{b},\vec{b'}$ drop after $k_2 \leq k_4$ and $2(k_2 + \log_2 q) - k_3 \leq k_4$, but at this point the protocol already produces incorrect results.}}
	\label{fig_implementation}
\end{figure}

\CHANGED{
We also evaluate the protocol's correctness as well as the effectiveness of our attacks depending on varying parameters. The results are presented in Figure~\ref{fig_implementation} for varying values of~$k_4$ and confirm the contradiction between the protocol's correctness and its security. Under the correctness constraints of Equations~\ref{eq:constraints4},~\ref{eq:constraints5},~\ref{eq:constraints6}, all of our attacks are close to 100\% accurate. Conversely, the protocol is entirely correct for these parameter choices as well. When Equation~\ref{eq:constraints5} is broken with $k_4 \geq k_2 - \log_2 q -\log_2 n$, the absolute correctness error starts to appear and rises rapidly. Shortly after this, when the accuracy threshold $k_4 \geq k_2$ of our first attack distinguishing any $\vec{b}$ is passed, its accuracy quickly drops to the baseline of~50\% (the accuracy of randomly guessing between two~$\vec{b}$). The same occurs after the threshold for our second distinguishing attack is passed, at which point the maximum correctness error of $k_1 - 2k_2 = 112$ bit is already reached. 
Furthermore, to demonstrate that our attacks even allow to test for a certain $\vec{b}$, we also evaluate both attacks by distinguishing a random $\vec{b}$ from a $\vec{b'}$ that only differs from~$\vec{b}$ by 1 in one position. Our evaluation shows that, though the accuracies drop earlier than for random $\vec{b'}$, these attacks work for the standard parameters and that therefore a precise testing and searching for $P_1$'s input is possible. Notably, we evaluate the correctness for $\vec{b}=\vec{0}$, as Equation~\ref{eq:constraints5} comes from the random addends resulting from all $b_i=0$. When using a completely random~$\vec{b}$, the correctness error only starts to appear at $k_4 \geq k_2 + 64 = 264$ but, since the protocol should be correct for any input, we display the results for $\vec{b}=\vec{0}$.

Our implementation establishes that in any application using the protocol,~$P_0$ can check whether~$P_1$ has a certain input (like, e.g., a certain illness in a healthcare application). Clearly, this is a severe violation of privacy and serves as a reminder that the security notions used by the protocol's security analysis~(cf.~\S\ref{sec:flaws:correctness}) are insufficient and that the established definitions based on indistinguishability~(cf.~\S\ref{sec:seccomp:formulation}) should be used instead.} As outlined in \CHANGED{\S\ref{sec:seccomp:impossiblity}}, similar attacks will inadvertently still be possible even if additional randomizations are introduced to prevent \CHANGED{these concrete attacks} as long as no cryptographic assumptions are utilized.

\section{Conclusion}
We showed in \CHANGED{\S\ref{sec:seccomp:impossiblity}} that protocols for \CHANGED{the }secure \CHANGED{two-party }computation of the scalar product imply oblivious transfer. As a result, such protocols very likely require public-key cryptography. Lu et al.'s protocol \cite{lu2013spoc,lu2014toward} is an example in academic use today that does not rely on such assumptions and is thus inherently insecure. Indeed, we found specific attacks that we have verified with an implementation\CHANGED{, showing that their protocol does not guarantee privacy}. With this comment paper we want to stress that (at least some) expensive public-key cryptography is necessary for such protocols and that new protocols should be proven secure \CHANGED{in established formal frameworks} to catch \CHANGED{such }flaws.


%

\ifCLASSOPTIONcompsoc
  \section*{Acknowledgments}
\else
\fi

We want to thank the anonymous reviewers for their helpful feedback. This work was supported by the DFG as part of project A.1 within the RTG 2050 ``Privacy and Trust for Mobile Users'' and as part of project E4 within the CRC 1119 CROSSING, and by the BMBF and the HMWK within CRISP.

\ifCLASSOPTIONcaptionsoff
  \newpage
\fi



\bibliographystyle{IEEEtran}
\bibliography{bibliography}
%

%








\end{document}